\documentclass[%
 reprint,
 amsmath,amssymb,
 aps,
prstab,
]{revtex4-1}

\usepackage{graphicx}
\usepackage{dcolumn}
\usepackage{bm}
\usepackage{gensymb}
\usepackage{color}
\usepackage{subcaption}

\begin{document}


\title{Electron Diffusion in Microbunched Electron Cooling}

\author{W. F. Bergan}
\email[]{wbergan@bnl.gov}
\affiliation{Brookhaven National Laboratory, Upton, New York 11973, USA}
\begin{abstract}

Coherent electron cooling is a novel method to cool dense hadron beams on timescales of a few hours. This method uses a copropagating beam of electrons to pick up the density fluctuations within the hadron beam in one straight section and then provide corrective energy kicks to the hadrons in a downstream straight, cooling the beam. Microbunched electron cooling is an extension of this idea which induces a microbunching instability in the electron beam as it travels between the two straights, amplifying the signal. However, initial noise in the electron bunch will also be amplified, providing random kicks to the hadrons downstream which tend to increase their emittance. In this paper, we develop an analytic estimate of the effect of the electron noise and benchmark it against simulations. We also discuss how this effect has impacted the cooler design.

\end{abstract}

\maketitle

\section{Introduction}\label{sec:intro}

\begin{figure}[!htbp]
\begin{center}
\includegraphics[width=1.0\columnwidth]{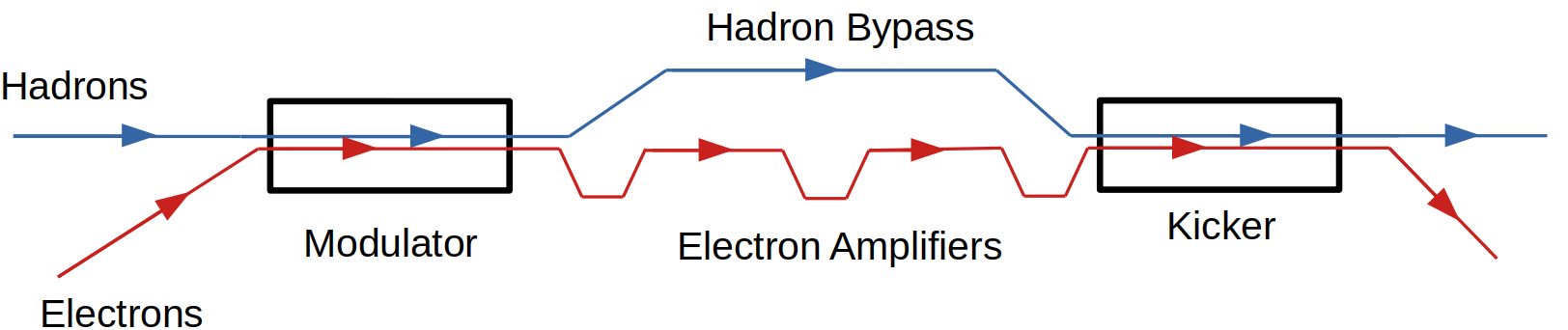}
\end{center}
\caption{\label{fig:setup} Schematic of the MBEC layout.}
\end{figure}

\begin{figure*}
\centering
        \centering
        \begin{tabular}{cc}

        \begin{subfigure}[b]{0.45\textwidth}
        \includegraphics[width=\textwidth]{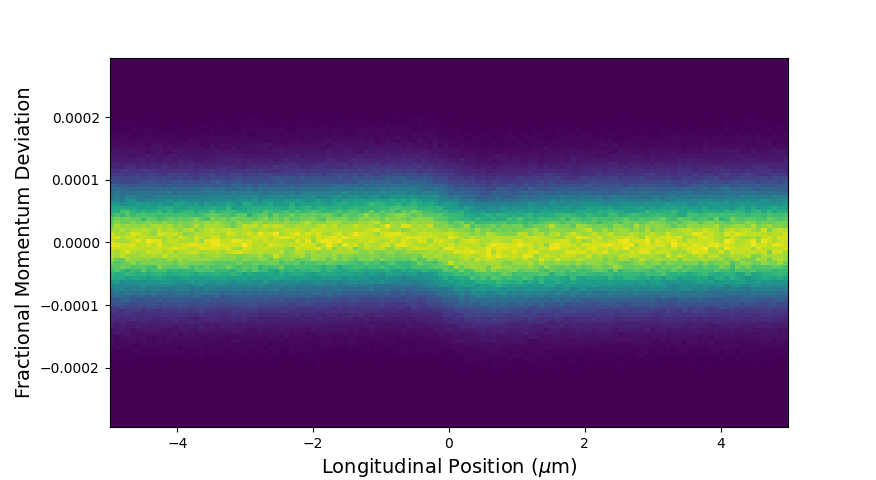}
        \caption{\label{subfig:start_p}}
        \end{subfigure}&

        \begin{subfigure}[b]{0.45\textwidth}
        \includegraphics[width=\textwidth]{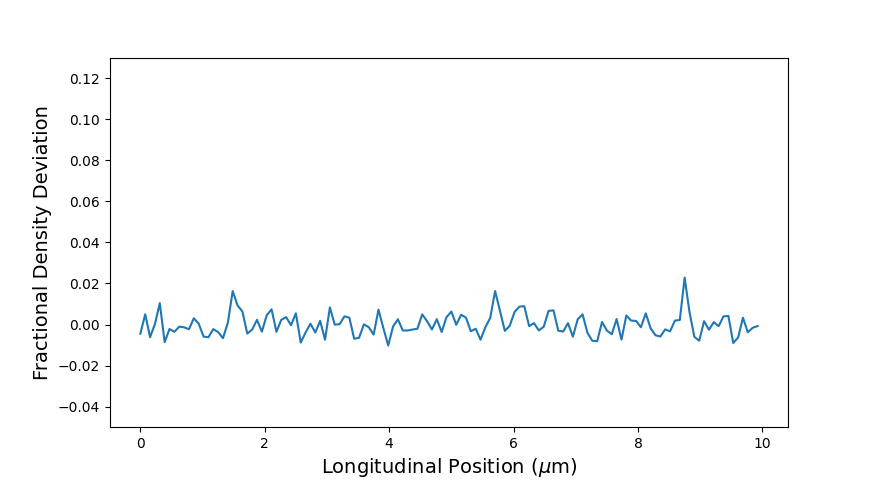}
        \caption{\label{subfig:start_d}}
        \end{subfigure}\\
        
        \begin{subfigure}[b]{0.45\textwidth}
        \includegraphics[width=\textwidth]{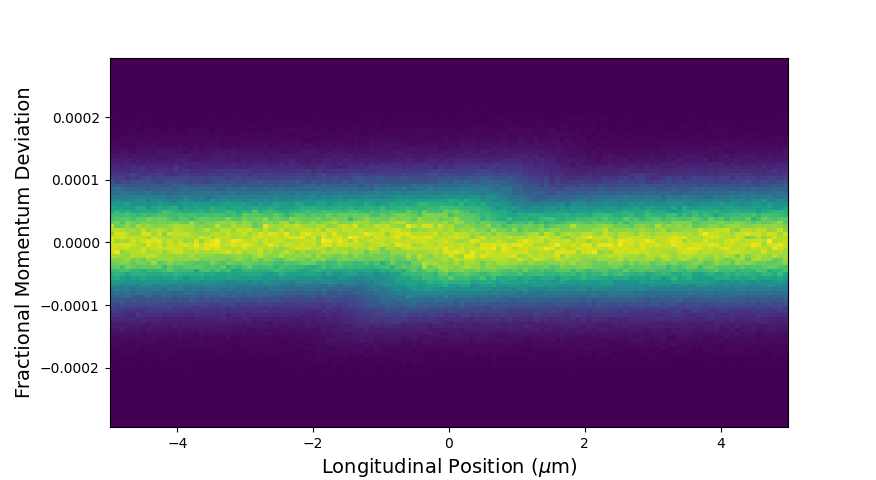}
        \caption{\label{subfig:0_p}}
        \end{subfigure}&

        \begin{subfigure}[b]{0.45\textwidth}
        \includegraphics[width=\textwidth]{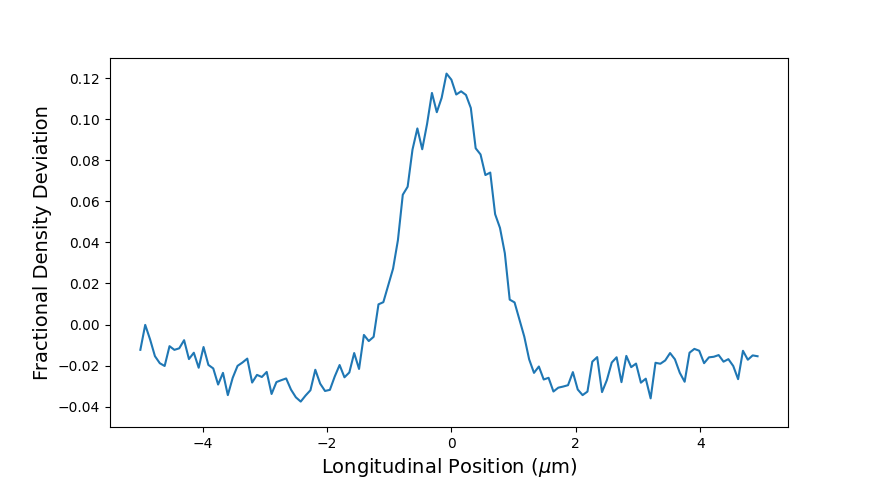}
        \caption{\label{subfig:0_d}}
        \end{subfigure}\\
        
        \begin{subfigure}[b]{0.45\textwidth}
        \includegraphics[width=\textwidth]{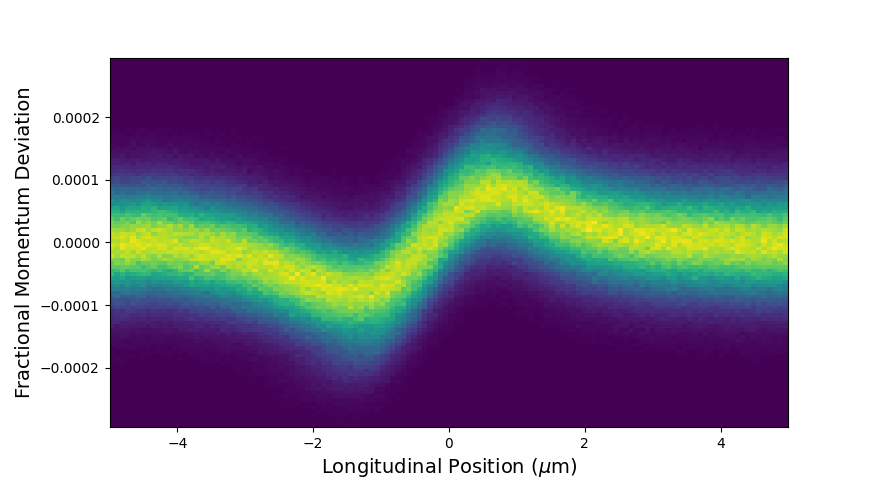}
        \caption{\label{subfig:3_p}}
        \end{subfigure}&

        \begin{subfigure}[b]{0.45\textwidth}
        \includegraphics[width=\textwidth]{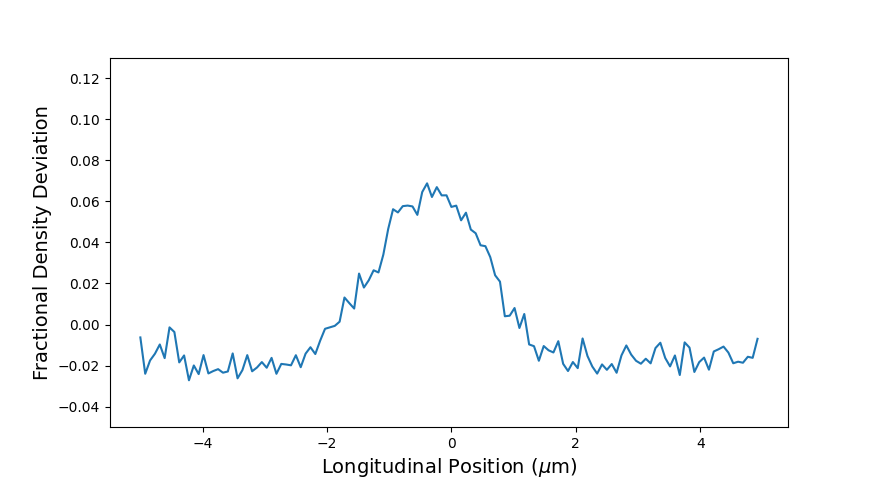}
        \caption{\label{subfig:3_d}}
        \end{subfigure}\\
        
        \begin{subfigure}[b]{0.45\textwidth}
        \includegraphics[width=\textwidth]{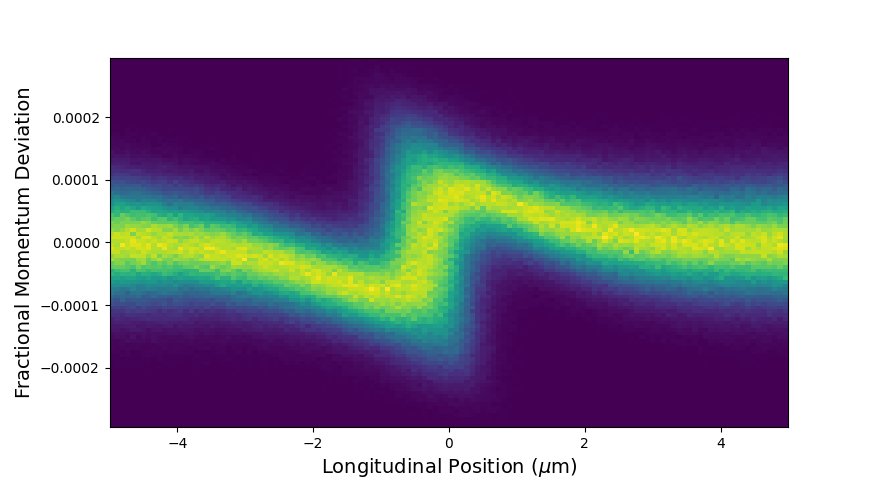}
        \caption{\label{subfig:end_p}}
        \end{subfigure}&

        \begin{subfigure}[b]{0.45\textwidth}
        \includegraphics[width=\textwidth]{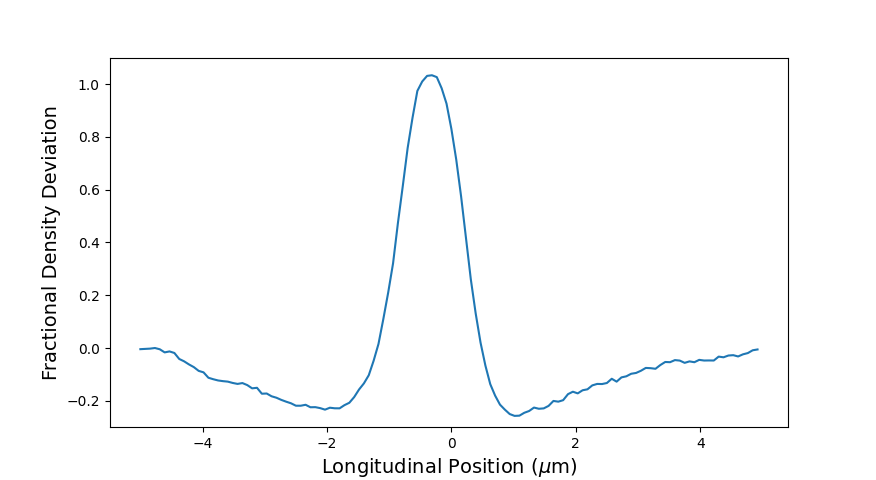}
        \caption{\label{subfig:end_d}}
        \end{subfigure}

        \end{tabular}
        
        \caption{\label{fig:amplification_time_series} Evolution of the 2D phase space of the electron beam through an amplification section, seeded by 18 thousand protons in the modulator with longitudinal positions near the origin. The initial energy perturbation from the hadron seen in subfig. \ref{subfig:start_p} and \ref{subfig:start_d} is rotated into a density perturbation after the first chicane, as seen in subfig. \ref{subfig:0_p} and \ref{subfig:0_d}. A partial plasma oscillation as the beam travels through the amplifier straight generates a significant energy modulation again, as in subfig. \ref{subfig:3_p} and \ref{subfig:3_d}, and the next chicane rotates this into an amplified density perturbation, as in subfig. \ref{subfig:end_p} and \ref{subfig:end_d}. Note the change in vertical scale in subfig. \ref{subfig:end_d} relative to the preceding plots.}
\end{figure*}

Microbunched electron cooling (MBEC) is a promising technique to cool dense hadron beams at timescales of a few hours, which will be necessary to prevent emittance blowup due to intrabeam scattering (IBS) at the planned electron-ion collider \cite{cite:eic_cdr}. This idea was introduced in \cite{cite:ratner} and the theory extensively developed in \cite{cite:stupakov_initial, cite:stupakov_amplifier, cite:stupakov_transverse, cite:stupakov_fourth}. A diagram of the setup is shown in Fig. \ref{fig:setup}. The basic concept is that the hadron beam to be cooled is propagated through a straight ``modulator'' section alongside an electron beam with the same relativistic gamma. During this time, a given hadron will provide energy kicks to the nearby electrons. The two beams are then separated, and the electron beam travels through a series of chicanes and straight sections, which use the microbunching instability to amplify the initially seeded energy perturbations and turn them into density fluctuations. Within the chicanes, the energy modulation of the electron beam is turned into a density fluctuation, with the scaling determined by the chicane's $R_{56}$. Within the straights, the electron beam undergoes roughly one quarter wavelength of a longitudinal plasma oscillation, so that these density fluctuations are turned back into energy fluctuations in preparation for passage through the next chicane. An illustration of the process is shown in Fig. \ref{fig:amplification_time_series}. Meanwhile, the transit time of the hadrons as they travel from modulator to kicker is dependent on their initial offsets in energy, transverse positions, and transverse angles. The electrons and hadrons then copropagate within a straight ``kicker'' section, where the density fluctuations in the electron beam provide energy kicks to the hadrons, with the dependence of the energy kick received by the hadron on its longitudinal delay in travelling from modulator to kicker termed the ``wake function.'' By correctly tuning the hadron modulator-to-kicker transfer matrix and the transverse dispersion and dispersion derivatives in the kicker, it can be arranged so that these energy kicks tend to decrease the initial transverse and longitudinal actions of the hadrons, providing a cooling force.

An important consideration in the operation of a cooler is diffusion; a given hadron sees not only its own wake, but also the wakes due to its neighbors, which provide a heating term partially counteracting the cooling. This has been previously discussed in \cite{cite:stupakov_initial, cite:stupakov_amplifier, cite:stupakov_transverse, cite:stupakov_fourth}. However, there is also a diffusion term originating in the noise in the electron beam; it will start off with initial energy and density fluctuations due to the Poisson statistics of discrete electrons and potentially some upstream instability, which will also be amplified within the cooler and provide energy kicks to the hadrons in the kicker. While \cite{cite:stupakov_amplifier} provides some discussion of the electron diffusion, this ignores the effect of the electron-electron energy kicks in the modulator. \cite{cite:eic_mbec_design} provides formulas related to this effect, but left their derivation vague, and provided no evidence of their accuracy. While direct particle-in-cell simulations, such as those discussed in \cite{cite:ipac2021_pic}, allow for direct computation of electron noise, such methods are time-consuming, rendering an analytic model invaluable for fast optimization and design work.


In Section \ref{sec:thry} of this paper, we show an explicit derivation of the electron diffusion. In Section \ref{sec:sim}, we compare the results of the theory to simulation.
In Section \ref{sec:design}, we discuss how the electron diffusion affects the cooler design and in Section \ref{sec:conclude} we present the conclusions.

\section{Theory}\label{sec:thry}

\subsection{Electron Wakes}\label{subsec:wakes}

In order to determine the diffusion due to noise in the electron beam, it is first necessary to determine the effect which a single electron in the modulator will have on the energy kick received by a hadron in the kicker, which we can think of as the ``electron wake.'' The process of determining the single electron wake is complicated by the fact that it enters into the cooling process in two distinct ways. First, any given electron in the modulator will provide energy kicks to nearby electrons, inducing an energy fluctuation in a manner almost exactly the same as a hadron does. Second, this electron will also propagate through the amplification sections itself; even if there were no modulator, the initial electron density modulation from shot noise alone would be amplified the same as one which had been induced due to energy kicks in the modulator.

We follow the approach of \cite{cite:stupakov_initial, cite:stupakov_amplifier} and use the conventions developed therein. We model the electrons and hadrons as rigid charged discs, with the charge density falling off as Gaussians in the transverse directions whose standard deviations equal the horizontal or vertical beam sizes of the electron or hadron beams, as appropriate. We also assume a perfectly linear wake. Since the typical wake length scale of a few microns is much shorter than the few millimeter or few centimeter electron and hadron bunch lengths, we will assume the currents in both bunches are constant over any regions of interest. We note that the amplification process detailed in \cite{cite:stupakov_amplifier} depends solely on the electron density modulations at the start of the first amplifier, after the first electron chicane. We will therefore end our analysis at this point, and use that paper's machinery to get the kick to the hadrons in the kicker. The effect of the electron noise therefore only enters in determining the longitudinal electron density perturbation at the start of the first amplifier straight, $\delta n_{(a)}(z)$.

Let the electron distribution entering the modulator be described by the longitudinal phase-space density
\begin{align}
        f_{(m,e)}(z,\eta) = n_eF_0 (\eta) + \delta f_{(m,e)}(z,\eta)
\end{align}
\noindent
where $z$ is the longitudinal position within the bunch, $\eta$ is the electron's fractional energy offset, $n_e$ is the constant component of the longitudinal electron density, $F_0(\eta)$ is the baseline electron energy distribution, and $\delta f_{(m,e)}$ is some perturbation to the above. Letting $\Delta\eta(z)$ be the energy kick an electron receives as a function of its longitudinal coordinate in the modulator, we may write the phase-space density at the start of the first amplifier straight, after passing through a chicane of strength $R_1$, as
\begin{align}
        f_{(a)}(z,\eta) &= f_{(m,e)}(z-R_1\eta, \eta - \Delta\eta(z - R_1\eta))\\\nonumber
        &=n_eF_0(\eta - \Delta\eta(z - R_1\eta))\\\nonumber
        &+ \delta f_{(m,e)}(z-R_1\eta, \eta - \Delta\eta(z - R_1\eta))
\end{align}

Subtracting off the background $n_eF_0 (\eta)$, Taylor expanding the first term to $1^{st}$ order in $\Delta\eta$, and keeping only the $0^{th}$ order of the second term (since it is already assumed to be a small perturbation), we arrive at
\begin{align}
        \delta f_{(a)}(z,\eta) &\approx -n_eF'_0(\eta)\Delta\eta(z - R_1\eta)\\\nonumber
        &+ \delta f_{(m,e)}(z-R_1\eta, \eta)
\end{align}
\noindent
and a corresponding frequency-space longitudinal density
\begin{align}
        \delta \hat{n}_{(a)}(k) &\equiv \int_{-\infty}^{\infty}dz e^{-ikz} \delta n_{(a)}(z)\\\nonumber
        &= \int_{-\infty}^{\infty} d\eta \int_{-\infty}^{\infty} dz e^{-ikz}\delta f_{(a)}(z,\eta)\\\nonumber
        &= \int_{-\infty}^{\infty} d\eta \int_{-\infty}^{\infty}dz e^{-ikz}[-n_eF'_0(\eta)\Delta\eta(z - R_1\eta)\\\nonumber
        &+ \delta f_{(m,e)}(z-R_1\eta, \eta)]\\\nonumber
        & = -\int_{-\infty}^{\infty} d\eta e^{-ikR_1\eta} n_eF'_0(\eta)\Delta\hat{\eta}(k)\\\nonumber
        &+ \int_{-\infty}^{\infty} d\eta \delta \hat{f}_{(a,\Delta\eta=0)}(k, \eta)
\end{align}
\noindent
where we have defined appropriate frequency-space density functions. We have also defined $\delta f_{(a,\Delta\eta=0)}(z, \eta) \equiv f_{(m,e)}(z-R_1\eta, \eta)$ as the longitudinal phase-space density of the electron beam at the start of the first amplifier straight if there were no energy kicks in the modulator.

If we assume that $F_0(\eta)$ describes a Gaussian energy distribution with RMS fractional energy spread $\sigma_{\eta}$, we arrive at
\begin{align}\label{eqtn:delta_eta_undef}
        \delta \hat{n}_{(a)}(k) &= -ikR_1\int_{-\infty}^{\infty} d\eta e^{-ikR_1\eta}n_eF_0(\eta)\Delta\hat{\eta}(k)\\\nonumber
        &+ \delta \hat{n}_{(a,\Delta\eta=0)}(k)\\\nonumber
        &= -ikR_1n_ee^{-k^2R_1^2\sigma_{\eta}^2/2}\Delta\hat{\eta}(k) + \delta \hat{n}_{(a,\Delta\eta=0)}(k)
\end{align}

We now need to calculate $\Delta\hat{\eta}(k)$. For the kick to the electrons due to the hadrons, this is provided by Eq. 54-56 of \cite{cite:stupakov_initial}:
\begin{align}\label{eqtn:hadr_kicks}
        \Delta\hat{\eta}_{(eh)}(k) = \frac{2iZr_eL_m}{\gamma^2\Sigma}H_{(m,eh)}(\varkappa)\delta\hat{n}_{(m,h)}(k)
\end{align}
with
\begin{align}\label{eqtn:H_definition}  
        H_{(m,eh)}(\varkappa) \equiv \int_0^{\infty} d\xi\Phi_{(m,eh)}(\xi)\sin(\varkappa\xi)
\end{align}
and
\begin{align}\label{eqtn:varkappa_definition}
        \varkappa \equiv \frac{k\Sigma}{\gamma}
\end{align}

$\Sigma$ is just some normalization length scale, taken equal to the horizontal hadron beam size in the modulator. However, it cancels out everywhere, and can be changed with impunity. The other variables used are $Z$, the hadron charge; $r_e$, the classical electron radius; $L_m$, the modulator length; $\gamma$, the relativistic gamma factor; and $\delta\hat{n}_{(m,h)}(k)$, the frequency-space hadron density perturbation in the modulator. $\Phi_{(m,eh)}(\xi)$ is the electron-hadron interaction function in the modulator, defined based on Appendix C of \cite{cite:stupakov_transverse} as:
\begin{align}\label{eqtn:phi_def}
        &\Phi_{(m,eh)}(\xi) = \frac{4\xi}{\sqrt{\pi}}\int_0^{\infty}d\lambda \lambda^2\\\nonumber
        &\times\frac{e^{-\lambda^2\xi^2}}{\sqrt{1+2\lambda^2(\Sigma_{x,h}^2 + \Sigma_{x,e}^2)/\Sigma^2}\sqrt{1+2\lambda^2(\Sigma_{y,h}^2 +\Sigma_{y,e}^2)/\Sigma^2}}
\end{align}
\noindent
where the various $\Sigma$ terms are the horizontal and vertical beam sizes of the hadrons and electrons, labelled appropriately.

We may insert Eq. \ref{eqtn:phi_def} into the definition of $H_{(m,eh)}(\varkappa)$ in Eq. \ref{eqtn:H_definition}. Using the antisymmetry of $\Phi$ to extend the $\xi$ integral from $-\infty$ to $\infty$ and doing it analytically, we obtain
\begin{align}\label{eqtn:h_def}
        &H_{(m,eh)}(\varkappa) = \int_0^{\infty}d\lambda \frac{\varkappa}{\lambda}\\\nonumber
        &\times\frac{e^{-\varkappa^2/4\lambda^2}}{\sqrt{1+2\lambda^2(\Sigma_{x,h}^2 + \Sigma_{x,e}^2)/\Sigma^2}\sqrt{1+2\lambda^2(\Sigma_{y,h}^2 +\Sigma_{y,e}^2)/\Sigma^2}}
\end{align}

For the kicks due to electron beam noise, we recognize that the physics is exactly the same, and so immediately write down
\begin{align}\label{eqtn:elec_kicks}
        \Delta\hat{\eta}_{(ee)}(k) &= -\frac{2ir_eL_m}{\gamma^2\Sigma}H_{(m,ee)}(\varkappa)\delta\hat{n}_{(m,e)}(k)
\end{align}

The changes from Eq. \ref{eqtn:hadr_kicks} are division by $-Z$ (the ratio of the hadron to electron charges), replacement of the hadron linear density by that of the electrons, and replacement of the electron-hadron $H$ function by the electron-electron version, which uses electron instead of hadron beam sizes in Eq. \ref{eqtn:h_def}.

Putting Eq. \ref{eqtn:hadr_kicks} and \ref{eqtn:elec_kicks} into Eq. \ref{eqtn:delta_eta_undef}, we obtain
\begin{align}\label{eqtn:density_amp}
        \delta \hat{n}_{(a)}(k) = &\frac{2Zr_eL_m}{\gamma^2\Sigma} kR_1n_ee^{-k^2R_1^2\sigma_{\eta}^2/2}\\\nonumber
        &\times\bigg(H_{(m,eh)}(\varkappa)\delta\hat{n}_{(m,h)}(k)\\\nonumber
        &- \frac{1}{Z}H_{(m,ee)}(\varkappa)\delta\hat{n}_{(m,e)}(k)\bigg)\\\nonumber
        &+ \delta \hat{n}_{(a,\Delta\eta=0)}(k)
\end{align}

From \cite{cite:stupakov_amplifier}, we see that the electron density at the kicker is equal to the electron density at the start of the first amplifier multiplied by the gain factors for the two amplifiers. These factors are denoted here by $G_2$ and $G_3$, with the number corresponding to the labelling of the associated electron chicane, and an expression for them is provided by Eq. 26 of \cite{cite:stupakov_amplifier}. Finally, it is evident that Eq. \ref{eqtn:hadr_kicks} describes the kick of the electron beam on the hadrons in the kicker if we multiply it by the ratio of the masses $m_e/m_h$, and change $L_m \rightarrow L_k$ (the kicker length), $H_{(m,eh)} \rightarrow H_{(k,eh)}$ (the electron-hadron interaction function in the kicker), and $\delta\hat{n}_{(m,h)}(k) \rightarrow \delta\hat{n}_{(k,e)}(k)$ (the electron density perturbation in the kicker). We arrive at an energy kick to the hadrons in the kicker:

\begin{align}\label{eqtn:energy_kicks_kicker}
        \Delta\hat{\eta}_{(k,h)}(k) = -\frac{r_h c}{\gamma}&\big(Z_h(k)\delta\hat{n}_{(m,h)}(k)\\\nonumber
        &+ Z_{e,1}(k)\delta\hat{n}_{(m,e)}(k)\\\nonumber
        &+ Z_{e,2}(k)\delta\hat{n}_{(a,\Delta\eta=0)}(k)\big)\\\nonumber
\end{align}
\noindent
where
\begin{align}\label{eqtn:impedances}
        Z_h(k) &= \frac{-4iL_mL_kI_e}{I_Ac\gamma^3\Sigma^2\sigma_{\eta}}\varkappa q_1e^{-\varkappa^2q_1^2/2}\\\nonumber
        &\times G_2G_3H_{(m,eh)}(\varkappa)H_{(k,eh)}(\varkappa)\\\nonumber
        Z_{e,1}(k) &= \frac{4iL_mL_kI_e}{ZI_Ac\gamma^3\Sigma^2\sigma_{\eta}}\varkappa q_1e^{-\varkappa^2q_1^2/2}\\\nonumber
        &\times G_2G_3H_{(m,ee)}(\varkappa)H_{(k,eh)}(\varkappa)\\\nonumber
        Z_{e,2}(k) &= \frac{-2iL_k}{Zc\gamma\Sigma}G_2G_3H_{(k,eh)}(\varkappa)
\end{align}
\noindent
are the impedances for the hadrons, the self-induced electron density modulations, and the pre-existing electron density modulations, respectively, $r_h = \frac{Z^2e^2}{4\pi\epsilon_0m_hc^2}$ is the classical hadron radius, $I_A = \frac{ec}{r_e} \approx 17$kA is the Alfv\'{e}n current, and $q_1 \equiv \frac{R_1\sigma_{\eta}\gamma}{\Sigma}$

Note that three terms are summed here. The first corresponds to the wake due to the hadrons, which has been discussed extensively in \cite{cite:stupakov_initial, cite:stupakov_amplifier, cite:stupakov_transverse}. The second corresponds to the wake due to the electrons which operates on the same principles as the hadron wake - each electron provides an energy kick to neighboring electrons in the modulator, resulting in a perturbation to the electron density in the kicker which in turn causes an energy kick to the hadrons. The third term is due to the fact that, in the absence of energy kicks in the modulator, the initial shot noise in the electron beam will still be present at the start of the first amplifier, and so will be amplified and provide a kick to the hadrons in the kicker. This third term is covered in \cite{cite:stupakov_amplifier}. We need to keep two terms associated with the electrons because the electron shot noise enters into our calculations both by seeding energy perturbations in the modulator and in its ``raw'' form at the start of the first amplifier, forcing us to account for the unperturbed electron density modulations in both locations. To look at it another way, the hadrons can be approximated to leading order as stationary in the modulator, and so we only have to deal with their one dimensional distribution, while the electrons will change their longitudinal coordinate between the modulator and first amplifier due to their energy offsets, so that we have to consider the full 2-dimensional longitudinal phase space, necessitating the use of two wake functions.

\subsection{Diffusion}\label{subsec:diffusion}
We are now ready to calculate the diffusion, including the contribution from noise in the electron beam. For convenience, we move from frequency space to physical space, with the definition
\begin{align}\label{eqtn:wake_def}
        w(z) = -\frac{c}{2\pi}\int_{-\infty}^{\infty}Z(k)e^{ikz}dk
\end{align}
\noindent
so that
\begin{align}
        \Delta\eta_{(k,h)}(z) = \frac{r_h}{\gamma}\int_{-\infty}^{\infty}\big[&w_h(z-z')\delta n_{(m,h)}(z')\\\nonumber
        + &w_{e,1}(z-z')\delta n_{(m,e)}(z')\\\nonumber
        + &w_{e,2}(z-z')\delta n_{(a,\Delta\eta=0)}(z')\big]dz'
\end{align}

We assume that the electrons and hadrons entering the modulator have pure Poisson shot noise. Since the integral of each of the three wakes is zero, this implies that the average kick a hadron receives due to this noise is zero. It now remains to calculate the RMS kick strength. The average squared kick is given by
\begin{widetext}
\begin{align}\label{eqtn:variance}
        \Big\langle\Delta\eta_{(k,h)}^2\Big\rangle = \bigg(\frac{r_h}{\gamma}\bigg)^2\Bigg\langle\int_{-\infty}^{\infty}\int_{-\infty}^{\infty}dz'dz'' \big[&w_h(z-z')\delta n_{(m,h)}(z')w_h(z-z'')\delta n_{(m,h)}(z'')\\\nonumber
        + &w_{e,1}(z-z')\delta n_{(m,e)}(z')w_{e,1}(z-z'')\delta n_{(m,e)}(z'')\\\nonumber
        + &w_{e,2}(z-z')\delta n_{(a,\Delta\eta=0)}(z')w_{e,2}(z-z'')\delta n_{(a,\Delta\eta=0)}(z'')\\\nonumber
        + &2w_{e,1}(z-z')\delta n_{(m,e)}(z')w_{e,2}(z-z'')\delta n_{(a,\Delta\eta=0)}(z'')\big]\Bigg\rangle
\end{align}
\end{widetext}

We have ignored the hadron-electron cross-terms, since there is no correlation between those two beams. However, we do need to be careful about the cross-term from the two electron wakes. The longitudinal density perturbation for the particles in either beam can be written as
\begin{align}
        \delta n(z) = \sum_i \delta(z - z^{(i)}) - n_0
\end{align}
\noindent
where $n_0$ is the average particle density and $z^{(i)}$ is the position of the $i^{th}$ particle. Going forward, we will ignore the constant-background $-n_0$ term at the end, since the total integral of the wake is 0.

Putting this into the first term of Eq. \ref{eqtn:variance}, we obtain
\begin{align}
        \bigg(\frac{r_h}{\gamma}\bigg)^2\bigg\langle\sum_{i,j}w_h(z-z_m^{(i)})w_h(z-z_m^{(j)})\bigg\rangle
\end{align}
\noindent
where $z_m^{(i)}$ refers to the position of particle $i$ in the modulator.

Since we assume white noise in the hadron beam, only the diagonal terms survive. We are then left with the sum over all hadron positions of the average squared hadron wake function, which can be rewritten as the integral
\begin{align}
        \bigg(\frac{r_h}{\gamma}\bigg)^2 n_h\int_{-\infty}^{\infty}w^2_h(z') dz'
\end{align}
\noindent
where $n_h$ is the mean linear hadron density.

The second and third terms of Eq. \ref{eqtn:variance} can be simplified using an identical procedure. The fourth term requires some special care. Recall that $\delta n_{(m,e)}(z)$ is the density of electrons at the modulator, while $\delta n_{(a,\Delta\eta=0)}(z)$ is the electron longitudinal density at the first amplifier if we ignore the energy kicks in the modulator. If electron $i$ is located at position $z_m^{(i)}$ in the modulator, with fractional energy offset $\eta^{(i)}$, it will be at position $z_m^{(i)} + R_1 \eta^{(i)}$ at the start of the first amplifier in the absence of modulator energy kicks. Using the same delta function and diagonalization arguments as above, we arrive at a simplification of the fourth term
\begin{align}
        2\bigg(\frac{r_h}{\gamma}\bigg)^2\bigg\langle\sum_{i}w_{e,1}(z-z_m^{(i)})&w_{e,2}(z-z_m^{(i)} - R_1\eta^{(i)})\bigg\rangle\\\nonumber
        =2\bigg(\frac{r_h}{\gamma}\bigg)^2 n_e \int_{-\infty}^{\infty}dz'\int_{-\infty}^{\infty}d\eta&\frac{1}{\sqrt{2\pi}\sigma_{\eta}}e^{-\eta^2/2\sigma_{\eta}^2}\\\nonumber
        \times &w_{e,1}(z')w_{e,2}(z'-R_1\eta)
\end{align}
\noindent
where we have assumed a Gaussian fractional energy spread of RMS $\sigma_{\eta}$ and longitudinal electron bunch density of $n_e$.

For a wake scale of $\sim1\mu$m, a 6cm hadron beam of $6.9\times10^{10}$ protons, and a 1nC beam of electrons with bunch length of roughly 1cm, we expect on the order of several hundred thousand particles from each beam to contribute to the diffusion term. Using the central limit theorem, the diffusive kick which a proton receives each turn can be approximated as being drawn from a Gaussian distribution with a mean of $0$ and a variance given by
\begin{align}\label{eqtn:diffusion_full}
        \Big\langle\Delta\eta_{(k,h)}^2 \Big\rangle=\bigg(\frac{r_h}{\gamma}\bigg)^2\int_{-\infty}^{\infty}&dz'\bigg(n_h w^2_h(z')\\\nonumber
        + &n_e w^2_{e,1}(z') + n_e w^2_{e,2}(z')\\\nonumber
        + &2n_e\int_{-\infty}^{\infty} d\eta \frac{1}{\sqrt{2\pi}\sigma_{\eta}}e^{-\eta^2/2\sigma_{\eta}^2}\\\nonumber
        \times &w_{e,1}(z')w_{e,2}(z'-R_1\eta)\bigg)
\end{align}

This procedure for simulating the diffusive kicks as random numbers drawn from an appropriate distribution is similar to the methods used in \cite{cite:gang_wang} to model IBS in a simulation of coherent electron cooling and in \cite{cite:st_wang} to model diffusion in a simulation of optical stochastic cooling.

\section{Simulation}\label{sec:sim}

\subsection{Nearly-Linear Simulation}

In order to verify the above calculations, we turn to a one-dimensional particle-in-cell (PIC) simulation, described previously in \cite{cite:ipac2021_pic} and \cite{cite:schottky}. We use parameters based on those currently planned for cooling 275GeV protons at the EIC, shown in Tab. \ref{tab:param}. However, in order to avoid nonlinear effects, we use only one amplification section, ie, we place the kicker immediately after the second electron chicane. We also track through the modulator and kicker in a single step to avoid the impact of plasma oscillations in those regions, with the effective $R_{56}$ of the modulator length added to the $R_{56}$ of the first chicane when obtaining the theory wakes.

This code creates the same number of electrons within a 10$\mu$m section of the beam as would be expected from the peak electron bunch density. These have a uniform longitudinal distribution and Gaussian momentum distribution. We also generate $10^5$ proton macroparticles within the same region with uniform longitudinal and Gaussian momentum distributions, each having the charge of 6.8 real protons. Since noise adds in quadrature, this is statistically equivalent to having the full $4.6 \times 10^6$ protons expected within a 10$\mu$m length at the bunch center.

Particles are propagated through the cooler with a simple kick-drift model. The kick is calculated using the usual particle-in-cell (PIC) formalism with periodic boundary conditions, 128 longitudinal bins of length $\sim$78nm, and with the electron-electron and electron-proton forces making use of the disc-disc interaction functions described in \cite{cite:stupakov_initial, cite:stupakov_amplifier, cite:stupakov_transverse} and discussed in Eq. \ref{eqtn:hadr_kicks} - \ref{eqtn:elec_kicks} of this paper. We convert these to fractional momentum kicks by $\Delta \delta = \Delta \eta/\beta^2$, where $\delta$ is the particle's fractional momentum deviation and $\beta \approx 1$ is the relativistic beta function. The longitudinal coordinates are then updated as $z \rightarrow z + \Delta s\delta/\gamma^2$, where $\Delta s$ is the step length and $\gamma$ is the relativistic gamma function common to both particle species. The chicanes are modelled as simple point elements translating the particles' longitudinal coordinates by $z \rightarrow z + R_{56}\delta$.

\begin{table*}[!hbt]
   \centering
   \caption{Parameters for Longitudinal and Transverse Cooling}
   \begin{tabular}{lc}
           \textit{\textbf{Geometry}}                                           &                      \\
           Modulator Length (m)                                                 & 33                   \\ 
           Kicker Length (m)                                                    & 33                   \\ 
           Number of Amplifier Straights                                        & 2                    \\ 
           Amplifier Straight Lengths (m)                                       & 49                   \\ 
           \textit{\textbf{Proton Parameters}}                                  &                      \\
           Energy (GeV)                                                         & 275                  \\
           Protons per Bunch                                                    & $6.9\times10^{10}$   \\
           Proton Bunch Length (cm)                                             & 6                    \\ 
           Proton Fractional Energy Spread                                      & $6.8\times 10^{-4}$  \\ 
           Proton Emittance (x/y) (nm)                                          & 11.3 / 1             \\ 
           Horizontal/Vertical Proton Betas in Modulator (m)                    & 21.0 / 19.6          \\ 
           Horizontal/Vertical Proton Dispersion in Modulator (m)               & 0.002 / 0.067        \\ 
           Horizontal/Vertical Proton Dispersion Derivative in Modulator (m)    & 0.030 / -0.005       \\ 
           Horizontal/Vertical Proton Betas in Kicker (m)                       & 21 / 20              \\ 
           Horizontal/Vertical Proton Dispersion in Kicker (m)                  & 0.002 / 0.067        \\ 
           Horizontal/Vertical Proton Dispersion Derivative in Kicker (m)       & 0.030 / -0.005       \\ 
           Proton Horizontal/Vertical Phase Advance (rad)                       & 3.17 / 4.76          \\ 
           Proton $R_{56}$ between Centers of Modulator and Kicker (mm)         & 1.348                \\
           \textit{\textbf{Electron Parameters}}                                &                      \\
           Energy (MeV)                                                         & 150                  \\
           Electron Bunch Charge (nC)                                           & 1                    \\ 
           Electron Peak Current (A)                                            & 13                   \\ 
           Electron Fractional Slice Energy Spread                              & $5.9 \times 10^{-5}$ \\ 
           Electron Normalized Emittance (x/y) (mm-mrad)                        & 2.8 / 2.8            \\ 
           Horizontal/Vertical Electron Betas in Modulator (m)                  & 21.4 / 21.4          \\ 
           Horizontal/Vertical Electron Betas in Kicker (m)                     & 7.9 / 7.9            \\ 
           Horizontal/Vertical Electron Betas in Amplifier Straights (m)        & 4.9 / 4.9            \\ 
           $R_{56}$ in First Electron Chicane (mm)                              & 12.0                 \\ 
           $R_{56}$ in Second Electron Chicane (mm)                             & -6.7                 \\ 
           $R_{56}$ in Third Electron Chicane (mm)                              & -6.8                 \\ 
           \textit{\textbf{Cooling Times}}                                      &                      \\
           Horizontal/Vertical/Longitudinal IBS/Beam-Beam Times (hours)         & 2.0 / 5.0 / 2.9      \\ 
           Horizontal/Vertical/Longitudinal Cooling Times (hours)               & 0.9 / 1.9 / 1.2      \\ 
   \end{tabular}
   \label{tab:param}
\end{table*}

\begin{figure}[!htbp]
\begin{center}
\includegraphics[width=1.0\columnwidth]{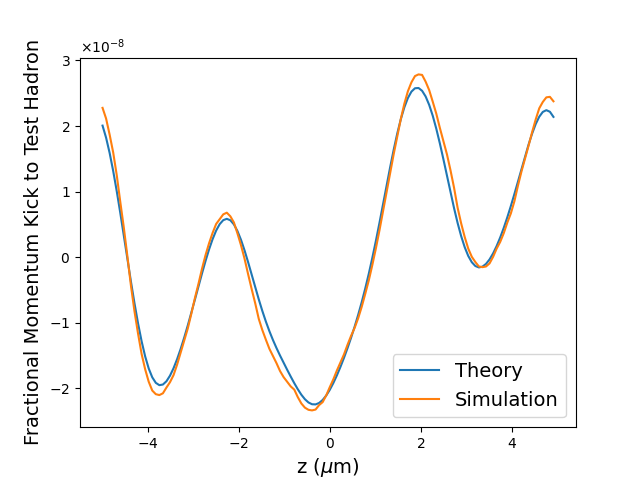}
\end{center}
\caption{\label{fig:all_noise} Fractional momentum kick to a proton in the kicker due to initial proton and electron noise in the modulator. Good agreement between theory and simulation is observed.}
\end{figure}

\begin{figure}[!htbp]
\begin{center}
\includegraphics[width=1.0\columnwidth]{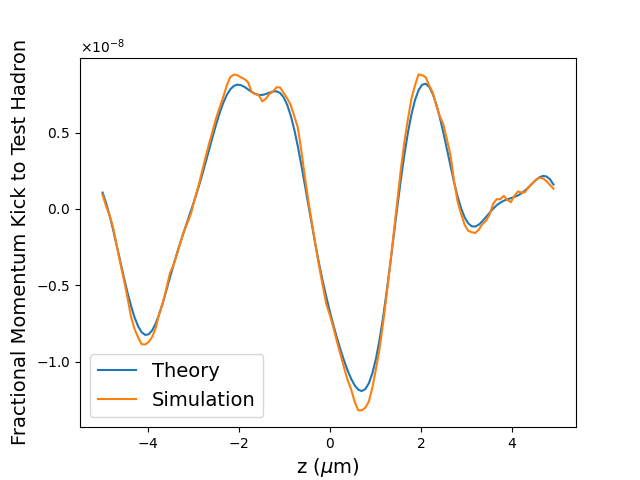}
\end{center}
\caption{\label{fig:elec_noise} Fractional momentum kick to a proton in the kicker due to only initial electron noise in the modulator. Good agreement between theory and simulation is observed.}
\end{figure}

In Fig. \ref{fig:all_noise}, we show the kick to a test proton in the kicker due to initial proton and electron noise in the modulator. We compare this to a theory curve made by convolving the initial proton and electron distributions with the wake functions derived from the impedances of Eq. \ref{eqtn:impedances}. We see that the theory reproduces the simulated result quite well. In order to focus on the electron wake, we rerun the simulation with no protons in the modulator, with the result shown in Fig. \ref{fig:elec_noise}. Again, good agreement with simulation is observed.

\begin{figure}[!htbp]
\begin{center}
\includegraphics[width=1.0\columnwidth]{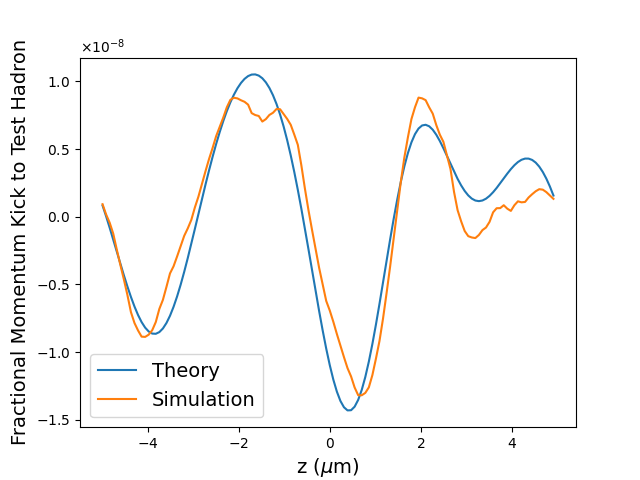}
\end{center}
\caption{\label{fig:elec_noise1} Fractional momentum kick to a proton in the kicker due to only initial electron noise in the modulator. The theory here only uses the middle impedance in Eq. \ref{eqtn:impedances}, and differs noticeably from the simulated result.}
\end{figure}

\begin{figure}[!htbp]
\begin{center}
\includegraphics[width=1.0\columnwidth]{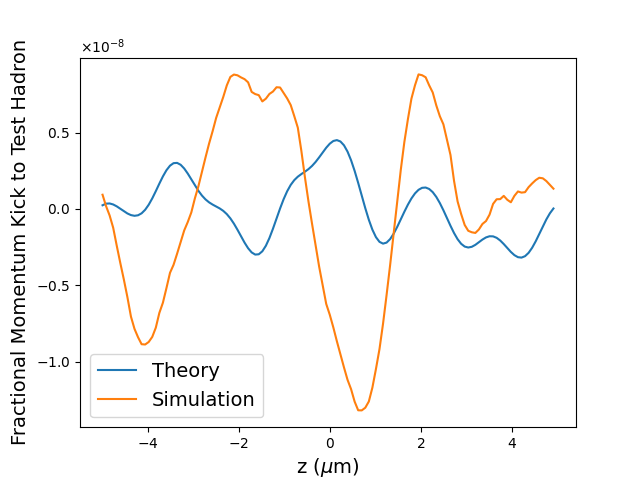}
\end{center}
\caption{\label{fig:elec_noise2} Fractional momentum kick to a proton in the kicker due to only initial electron noise in the modulator. The theory here only uses the final impedance in Eq. \ref{eqtn:impedances}, and is very different from the simulated result.}
\end{figure}

As a check of the necessity of both terms in the electron wake, we compute the electron theory curve using only the electron wakes corresponding to $Z_{e,1}(k)$ and $Z_{e,2}(k)$ individually, with the results shown in Fig. \ref{fig:elec_noise1} and \ref{fig:elec_noise2}, respectively. We see that using $Z_{e,1}(k)$ alone shows noticeable discrepancies from the simulated result, and using $Z_{e,2}(k)$ alone produces theory kicks which have very little correspondence to what is seen in simulation. Both terms are therefore necessary for properly understanding the evolution of the beam.

\begin{figure}[!htbp]
\begin{center}
\includegraphics[width=1.0\columnwidth]{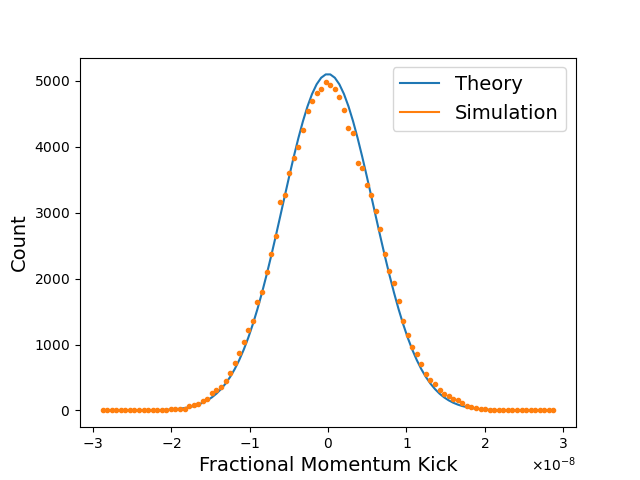}
\end{center}
\caption{\label{fig:histogram} Histogram of fractional momentum kicks to a proton in the kicker as obtained from the simulation and compared to the theoretical Gaussian distribution. Decent agreement is observed, although additional physics included in the simulation does lead to slight discrepancies.}
\end{figure}

We finally turn to the question of diffusion. We run the simulation 1000 times, starting with only electron noise, and record the fractional momentum kick which a test proton would receive in the kicker in each case at 128 locations over a 10$\mu$m length of beam. A histogram of the results is shown in Fig. \ref{fig:histogram}, along with the expected Gaussian distribution with the variance given by Eq. \ref{eqtn:diffusion_full}. The RMS kick received in simulation is $6.04 \times 10^{-9} \pm 0.05 \times 10^{-9}$, similar to, but slightly above, the value of $5.81\times 10^{-9}$ obtained from Eq. \ref{eqtn:diffusion_full}. Since the simulation includes additional physics, such as keeping the full longitudinal phase-space distribution at the start of the amplifier rather than projecting to the spatial plane, exact agreement is not expected.

If we recompute the theory value either ignoring the cross-term of Eq. \ref{eqtn:diffusion_full} or assuming that the two separate electron wakes can be simply added, we obtain RMS kicks of $7.77 \times 10^{-9}$ and $4.31 \times 10^{-9}$, respectively, showing that the form of the cross-term we have derived is both necessary and consistent with simulation. For comparison, the theoretical RMS fractional momentum kick from the proton noise is $11.3 \times 10^{-9}$, so that the electron contribution to diffusion cannot be neglected.

\subsection{Realistic Simulation}

In order to understand the impact which the electron noise will have in the actual cooler, we remove some of the approximations used for the nearly-linear simulation
described above. In particular, we include the full two amplifiers, leading to saturation effects in the electron beam; track through finite-length modulator and kicker sections, leading to plasma oscillations and variations in proton beam size in those elements; and add delays due to the transverse actions of the electrons. We can easily update the linear theory to handle to variations in the proton beam size in the modulator and kicker, but the other effects are more difficult to include in this manner. We discuss these effects and their impacts in more detail below.

\subsubsection{Two Amplifiers}

\begin{figure*}
\centering
        \centering
        \begin{tabular}{cc}

        \begin{subfigure}[b]{0.45\textwidth}
        \includegraphics[width=\textwidth]{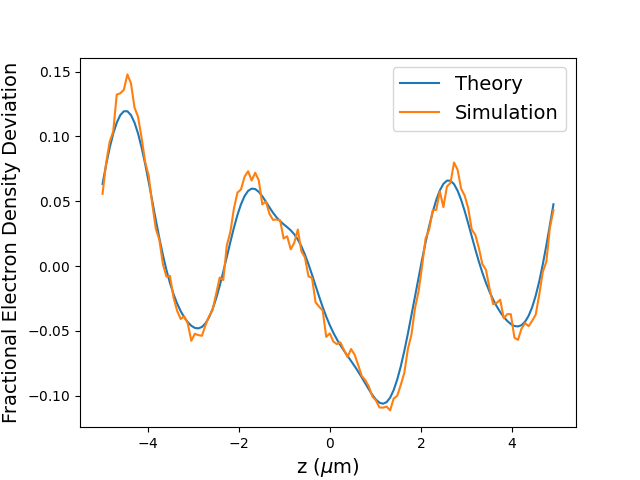}
        \caption{\label{subfig:sat_1amp} Density modulation with 1 amplifier.}
        \end{subfigure}&
        
        \begin{subfigure}[b]{0.45\textwidth}
        \includegraphics[width=\textwidth]{kick_vs_z_rev1.png}
        \caption{\label{subfig:kick_1amp} Momentum kick to protons with 1 amplifier.}
        \end{subfigure}\\

        \begin{subfigure}[b]{0.45\textwidth}
        \includegraphics[width=\textwidth]{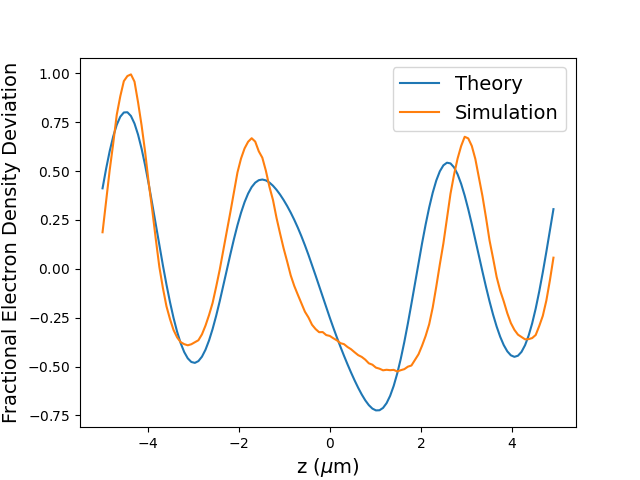}
        \caption{\label{subfig:sat_2amp} Density modulation with 2 amplifiers.}
        \end{subfigure}&
        
        \begin{subfigure}[b]{0.45\textwidth}
        \includegraphics[width=\textwidth]{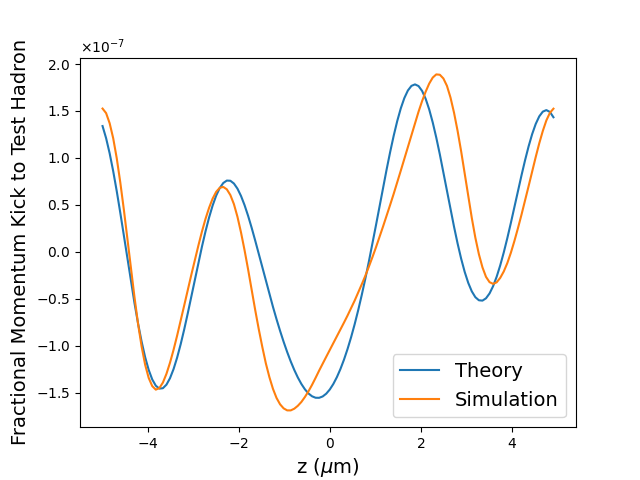}
        \caption{\label{subfig:kick_2amp} Momentum kick to protons with 2 amplifiers.}
        \end{subfigure}

        \end{tabular}
        
        \caption{\label{fig:add_saturation} Electron density modulation and fractional momentum kick to protons in the kicker when using either one amplifier or two. We see that, while with 1 amplifier we only have up to $\sim10\%$ fractional electron density modulations at the start of the kicker, 2 amplifiers brings these fractional modulations up to $\sim70\%$. As a consequence, we see disagreements with the linear theory.}
\end{figure*}

We simulate the full two amplifiers, rather than only using the first one. This additional amplification provides a larger wake for cooling, but also further amplifies the noise and increases the fractional density deviations in the electron beam. To the extent that these density deviations are a significant fraction of one, we approach the nonlinear saturated regime, as can be seen in Fig. \ref{fig:add_saturation}.

\subsubsection{Finite-Length Modulator and Kicker}

We also track the protons and electrons through the modulator and kicker in multiple steps, as opposed to the single-step tracking in the nearly-linear simulation. In using the more realistic multi-step tracking, the electrons will provide energy kicks to one another in the modulator and kicker, and their longitudinal positions will drift based on their energy offsets. The combination of these effects is the basis for plasma oscillations. A trivial extension of Eq. 17 of \cite{cite:stupakov_amplifier} gives the plasma oscillation wavelength at wavenumber $k$ as

\begin{align}
    \frac{1}{\lambda^2} = \frac{2kn_e r_e}{(2\pi)^2\Sigma\gamma^4\beta^2}H\bigg(\frac{k\Sigma}{\gamma}\bigg)
\end{align}

\noindent
where the function $H$ is either $H_{(m,ee)}$ or $H_{(k,ee)}$ in the modulator or kicker, respectively, and $\beta \approx 1$ is the relativistic beta.

With the parameters of Tab. \ref{tab:param}, we see that the oscillation wavelength is 564m in the modulator and 375m in the kicker at the average wake wavenumber (1.4$\mu$m$^{-1}$). The 33m modulator and kicker therefore represent only 6\% and 9\% of a plasma wavelength, respectively, so that we expect the plasma oscillations to have a small effect.

While we can use appropriate optics to maintain the electron beam size roughly constant over the length of each of the straight sections, the protons, having nearly 2000 times higher momentum, will not be significantly affected by the magnetic elements, and so will see the modulator and kicker straights as essentially drifts. The consequence is that the proton beam sizes, and therefore the proton-electron interaction functions, evolve over the course of these elements.

The nominal proton beta functions from Tab. \ref{tab:param} describe the beam at the modulator and kicker centers, with the optics at other places obtained through the evolution through a drift. These effects will be important if the proton beam size changes appreciably over the length of the modulator or kicker. For our parameters, the horizontal and vertical beam sizes increase by roughly 45\% and 40\%, respectively, in going from the center of the modulator or kicker straight to the ends, and so the beam size variation is not negligible. This effect is included in the simulation by using a new proton/electron interaction function at each step in these elements. It is also included in the theory calculation by computing the $H$ functions using the proton beam size at various locations in the modulator and kicker and taking the average value.

\begin{figure*}
\centering
        \centering
        \begin{tabular}{cc}

        \begin{subfigure}[b]{0.45\textwidth}
        \includegraphics[width=\textwidth]{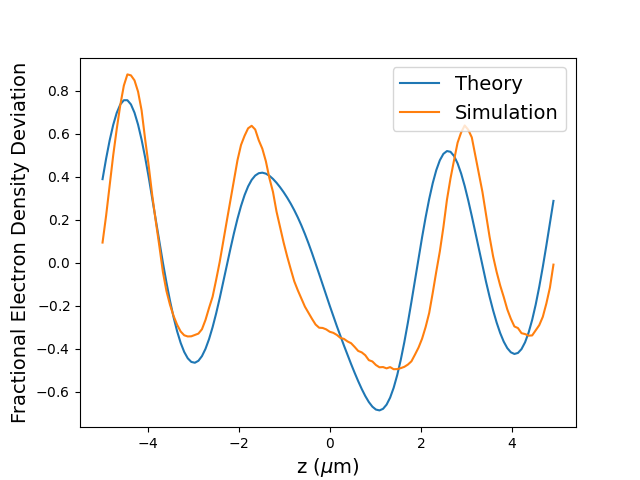}
        \caption{\label{subfig:sat_2amp_multi} Density modulation.}
        \end{subfigure}&
        
        \begin{subfigure}[b]{0.45\textwidth}
        \includegraphics[width=\textwidth]{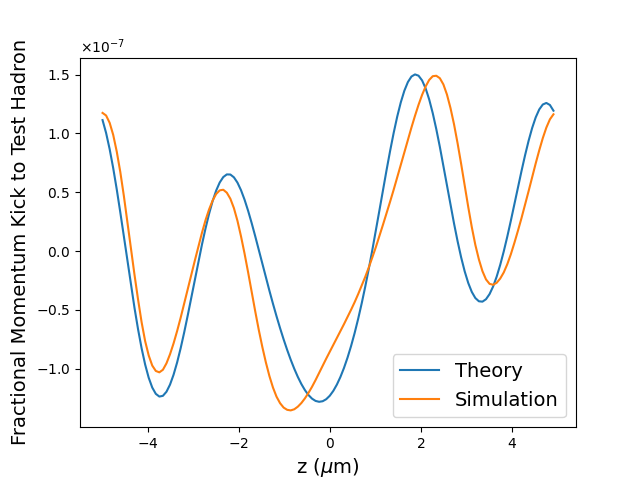}
        \caption{\label{subfig:kick_2amp_multi} Fractional momentum kick.}
        \end{subfigure}

        \end{tabular}
        
        \caption{\label{fig:finite_mk} Electron density modulation and fractional momentum kick to protons in the kicker when using two amplifiers and treating the modulator and kicker as elements of finite length. While both are somewhat decreased due to variations in the proton beam size, the agreement is similar to that shown in the 2-amplifier plots in Fig. \ref{fig:add_saturation}.}
\end{figure*}

A comparison of the theory and simulation with these modifications arising from finite modulator and kicker lengths is shown in Fig. \ref{fig:finite_mk}. We see that while there is a slight reduction in the amplitude in theory and simulation due to the larger proton beam sizes in the modulator and kicker, the agreement between the two is similar to what we had in the 2-amplifier plots of Fig. \ref{fig:add_saturation}.

\subsubsection{Transverse-Dependent Delays}

Up to now, we have considered only the longitudinal motion of the electrons and protons. However, they will move in all three dimensions during their passage through the cooler. One consequence of this is that electrons with large transverse actions gain additional longitudinal delays due to the contributions of the second-order transfer matrix elements. In particular, the path length which an electron with transverse angles $x'$ and $y'$ travels in moving a longitudinal distance $\Delta s$ is $\Delta s\sqrt{1 + x'^2 + y'^2}$. Taylor expanding and taking the betatron phase average, we find that an electron with transverse actions $J_x$ and $J_y$ travelling through a region with Courant-Snyder gamma functions $\gamma_x$ and $\gamma_y$ has an additional longitudinal shift

\begin{align}\label{eqtn:transverse_delay}
\Delta z = -\frac{\gamma_x J_x + \gamma_y J_y}{2} \Delta s
\end{align}

We include this effect in simulation by assigning each electron actions from the expected exponential distribution, and giving it an extra longitudinal shift at each time step as described in Eq. \ref{eqtn:transverse_delay}. We pick the gamma functions to be 3 divided by the beta function in the relevant plane in order to account for the fact that the Courant-Snyder alpha cannot always be equal to 0. See \cite{cite:ipac2024} for further details. This effect will be small if the delay described by Eq. \ref{eqtn:transverse_delay} is small relative to the wake wavelength when travelling between the modulator and kicker centers. The parameters in Tab. \ref{tab:param} give a typical delay of 0.7$\mu$m. This is non-negligible for our wake, which has its peak roughly 0.9$\mu$m from the zero-crossing.

\begin{figure*}
\centering
        \centering
        \begin{tabular}{cc}

        \begin{subfigure}[b]{0.45\textwidth}
        \includegraphics[width=\textwidth]{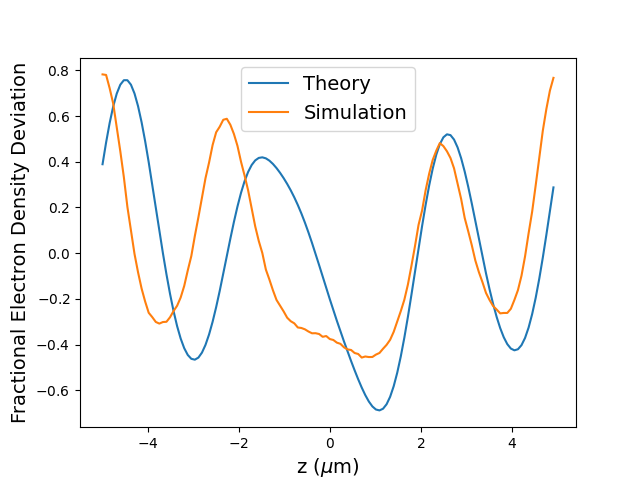}
        \caption{\label{subfig:sat_transverse} Density modulation.}
        \end{subfigure}&
        
        \begin{subfigure}[b]{0.45\textwidth}
        \includegraphics[width=\textwidth]{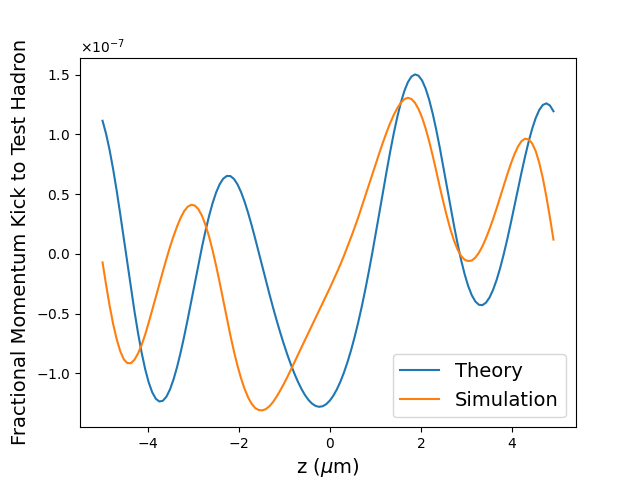}
        \caption{\label{subfig:kick_transverse} Fractional momentum kick.}
        \end{subfigure}

        \end{tabular}
        
        \caption{\label{fig:transverse} Electron density modulation and fractional momentum kick to protons in the kicker when using two amplifiers, treating the modulator and kicker as elements of finite length, and including additional delays due to the electrons' transverse motion. We see some reduction in the simulated values relative to Fig. \ref{fig:finite_mk}, as well as a prominent longitudinal shift.}
\end{figure*}

If we include this effect in the simulation as well, we obtain the curves shown in Fig. \ref{fig:transverse}. We see that the simulated saturation and kick amplitudes decrease somewhat relative to what we had in Fig. \ref{fig:finite_mk}. We also note a systematic shift towards negative $z$ values, since the extra transverse delay always causes electrons to arrive in the kicker later. This shift has no impact on the diffusion.

\subsubsection{Diffusion Calculation}

With all the above effects included, we can recalculate the RMS energy kicks provided by the noise in the electron and hadron beams. These are shown in Tab. \ref{tab:rms_kicks} for both theory and simulation \footnote{We do not simulate the case of using proton noise only, since it is not possible to run the simulation with a noiseless electron beam.}. We see that, as with the nearly-linear case, the contribution of the electrons to the diffusion is on the same scale as the proton contribution. Even with the addition of the nonlinear effects, the theory still holds up quite well relative to the simulation.

\begin{table*}[!hbt]
   \centering
   \caption{RMS Momentum Kick to Test Proton}
   \begin{tabular}{lcccccccccc}
                                                 &&&&&  \textbf{Theory}     &&&&& \textbf{Simulation}                           \\
                    Electron and Proton Noise    &&&&& $7.62 \times 10^{-8}$ &&&&& $7.26 \times 10^{-8} \pm 0.07 \times 10^{-8}$ \\
                    Proton Noise Only            &&&&& $6.65 \times 10^{-8}$ &&&&& Not Simulated                                 \\
                    Electron Noise Only          &&&&& $3.72 \times 10^{-8}$ &&&&& $3.54 \times 10^{-8} \pm 0.04 \times 10^{-8}$ \\
   \end{tabular}
   \label{tab:rms_kicks}
\end{table*}

\section{Practical Design Considerations}\label{sec:design}

The addition of electron noise will increase the level of saturation in the kicker. While we have not explicitly written the equation in this paper, it is relatively easy to obtain if we do not multiply by the electron-hadron interaction term when we go from Eq. \ref{eqtn:density_amp} to Eq. \ref{eqtn:energy_kicks_kicker}. Since the merit function used in our optimization procedure includes a penalty term for solutions with large saturation, adding the contribution of electrons to the saturation tends to reduce the ideal system gain from what the optimizer would have found if it had only considered the proton contribution. As an example, a 15\% higher total amplification gain would increase the wake amplitude by the same amount while giving the same saturation (assuming only proton noise) as we have from our current design parameters (using both proton and electron noise).

Analytic formulas for the contribution of the electrons to the saturation also allows us to set quantitative limits on the amount of noise above Poisson shot noise which we can tolerate in the electron beam. For example, if the electron and proton beams have only Poisson shot noise, the parameters of Tab. \ref{tab:param} give an RMS fractional density modulation of the electron beam in the kicker of 0.33. If we wish to limit this saturation to be no more than 10\% larger than its nominal value, we find that the noise power in the electron beam must be no more than 1.8 times its Poisson value.

There is another important effect to consider. A look at Eq. \ref{eqtn:impedances} shows that the relative sign of the two electron impedances depends on the sign of the $R_{56}$ in the first chicane. A negative $R_{56}$ leads to these two impedance terms having the same sign and adding constructively, while a positive $R_{56}$ leads to the two terms having opposite signs and partially cancelling. This is understood physically as the negative $R_{56}$ giving the electrons an effective negative mass, so that their space-charge repulsion will tend to make them clump together, while a positive $R_{56}$ will cause their mutual repulsion to push them apart. Naturally, the latter option is preferred in order to minimize the electron diffusion.

In a previous version of the cooler design, we had in fact chosen to use a negative $R_{56}$ in the first chicane in the hope that having the smaller negative chicanes first would help delay the onset of saturation and reduce the nonlinear effects. While there have been a number of changes in the lattice parameters between that optimization and those described in Tab. \ref{tab:param}, we can make an estimate of the effect of this change by swapping the first and third chicanes in the current parameters. While the linear formulas show that the wake function and proton diffusion are invariant under reordering of the chicanes, the electron diffusion is not. In particular, the RMS kick from noise in the electron beam increases from $3.72 \times 10^{-8}$ to $6.65 \times 10^{-8}$ according to the linear theory. Adding this in quadrature to the $6.65 \times 10^{-8}$ RMS kick from proton noise, we find that the total RMS momentum kick increases from $7.62 \times 10^{-8}$ to $9.40 \times 10^{-8}$. Similarly, RMS saturation in the electron beam at the kicker increases from 0.32 to 0.39.

\section{Conclusion}\label{sec:conclude}

We have derived formulas for the diffusion expected in a microbunched electron cooling system due to initial shot noise in the electron beam. Along the way, we have shown how to obtain the response to an arbitrary electron distribution in the modulator. For a reasonable set of parameters, this can be comparable to the proton noise. Properly understanding electron diffusion allows better estimates of the degree of saturation which will be induced by the time we reach the kicker, resulting in a lower optimal gain, as well as allowing us to set limits on initial electron beam noise levels. It also allows us to quantitatively understand the impact of the sign of the $R_{56}$ of the first electron chicane; picking it to be positive rather than negative has a significant impact in reducing the total diffusion and saturation.

\section{Acknowledgements}\label{sec:ack}

This work was supported by Brookhaven Science Associates, LLC under Contract No. DE-SC0012704 with the U.S. Department of Energy. I would also like to thank Erdong Wang, Michael Blaskiewicz, and Panagiotis Baxevanis for useful discussions.

\bibliography{elec_noise_prab}

\end{document}